\documentclass[
    reprint,
    superscriptaddress,
    preprintnumbers,
    amsmath,amssymb,
    aps,
    pra,
    10pt
]{revtex4-2}

\usepackage{graphicx}% Include figure files
\usepackage{dcolumn}% Align table columns on decimal point
\usepackage{bm}% bold math
\usepackage{color}% additionally defined
\usepackage{physics}
\usepackage{bm}
\usepackage[normalem]{ulem}% additionally defined

\usepackage[mathlines]{lineno}% Enable numbering of text and display math
\usepackage{comment}
\usepackage{quantikz}
\usepackage{rotating}
\begin{document}

\preprint{APS/QGFA}

\title{Quantum Gradient Flow Algorithm for Symmetric Positive Definite Systems \\ via Quantum Eigenvalue Transformation: Towards Quantum CAE} % Force line breaks with \\
%\thanks{A footnote to the article title}%
\author{Yuto Lewis Terashima}
    \email{yuto-terashima@aist.go.jp}
    \affiliation{
        National Institute of Advanced Industrial Science and Technology (AIST),\\
        Global R\&D Center for Business by Quantum--AI Technology (G--QuAT),\\
        Central 2, 1-1-1 Umezono, Tsukuba, Ibaraki 305-8568, Japan
    }%
    \affiliation{
        Keio University, Graduate School of Science and Technology,\\
        3-14-1 Hiyoshi, Kohoku--ku, Yokohama, Kanagawa 223-8522, Japan
    }%
\author{Tadashi Kadowaki}
    \affiliation{
        National Institute of Advanced Industrial Science and Technology (AIST),\\
        Global R\&D Center for Business by Quantum--AI Technology (G--QuAT),\\
        Central 2, 1-1-1 Umezono, Tsukuba, Ibaraki 305-8568, Japan
    }%
    \affiliation{DENSO CORPORATION, 1-1-4, Haneda Airport, Ota-ku, Tokyo 144-0041, Japan}
\author{Yohichi Suzuki}
    \affiliation{
        National Institute of Advanced Industrial Science and Technology (AIST),\\
        Global R\&D Center for Business by Quantum--AI Technology (G--QuAT),\\
        Central 2, 1-1-1 Umezono, Tsukuba, Ibaraki 305-8568, Japan
    }%
% \author{Mayu Muramatsu}
%     \affiliation{
%         Keio University, Department of Mechanical Engineering,\\
%         3-14-1 Hiyoshi, Kohoku-ku, Yokohama, Kanagawa 223-8522, Japan
%     }%
\author{Katsuhiro Endo}
 \email{katsuhiro.endo@aist.go.jp}
    \affiliation{
        National Institute of Advanced Industrial Science and Technology (AIST),\\
        Materials DX Research Center,\\
        Central 2, 1-1-1 Umezono, Tsukuba, Ibaraki 305-8568, Japan
    }%
    \affiliation{
        National Institute of Advanced Industrial Science and Technology (AIST),\\
        Global R\&D Center for Business by Quantum--AI Technology (G--QuAT),\\
        Central 2, 1-1-1 Umezono, Tsukuba, Ibaraki 305-8568, Japan
    }%

\date{\today}% It is always \today, today,
             %  but any date may be explicitly specified

\begin{abstract}
    In this study, we propose the Quantum Gradient Flow Algorithm (QGFA), a novel quantum algorithm for solving symmetric positive definite (SPD) linear systems based on the variational formulation and time--evolution dynamics.
    Conventional quantum linear solvers, such as the quantum matrix inverse algorithm (QMIA), focus on approximating the matrix inverse through quantum signal processing (QSP).
    However, QMIA suffers from a crucial drawback: its computational efficiency deteriorates as the condition number increases.
    In contrast, classical SPD linear solvers, such as the steepest descent and conjugate gradient methods, are known for their fast convergence, which stems from the variational optimization principle of SPD systems.
    Inspired by this, we develop QGFA, which obtains the solution vector through the gradient--flow process of the corresponding quadratic energy functional.  
    To validate the proposed method, we apply QGFA to the displacement--based finite element method (FEM) for two--dimensional linear elastic problems under plane stress conditions. 
    The algorithm demonstrates accurate convergence toward classical FEM solutions even with a moderate number of QSP phase factors.
    Compared with QMIA, QGFA achieves lower relative errors and faster convergence when initialized with suitable initial states, demonstrating its potential as an efficient preconditioned quantum linear solver.
    The proposed framework provides a physically interpretable connection between classical iterative solvers and quantum computational paradigms.
    These findings suggest that QGFA can serve as a foundation for future developments in Quantum Computer--Aided Engineering (Quantum CAE), including nonlinear and multiphysics simulations.
\end{abstract}

% \keywords{QSVT, Quantum CAE}%Use showkeys class option if keyword
                              %display desired
\maketitle
%\tableofcontents

\section{Introduction}\label{sec:intro}

The year 2025 has been designated by the United Nations as the International Year of Quantum Science and Technology to promote global awareness of the progress and potential of quantum technologies.  
This designation also commemorates the 100th anniversary of Heisenberg’s pioneering work, which laid the foundation of modern quantum mechanics~\cite{heisenberg1985quantentheoretische}.  
Among the various advances in quantum science, quantum computing has emerged as one of the most promising technologies.  
Quantum computers exploit the principles of superposition and entanglement, providing fundamentally new computational resources that far surpass those of classical computers.  
However, quantum systems based on superposition are intrinsically restricted to linear operations, posing a major challenge in realizing nonlinear transformations that are essential for many classical algorithms.  
Recent developments in algorithmic frameworks have provided powerful tools to overcome this limitation.
The Quantum Singular Value Transformation(QSVT)~\cite{martyn2021grand,gilyen2019quantum} has established a unified theoretical foundation for quantum algorithm design.  
By combining Quantum Signal Processing (QSP)~\cite{motlagh2024generalized,dong2021efficient,low2017optimal} and qubitization~\cite{low2019hamiltonian}, it has become possible to implement a wide class of effectively nonlinear matrix functions within the subspaces of unitary quantum systems.
QSVT has been further extended to diverse applications, including nonlinear operator transformations~\cite{guo2024nonlinear}, and to the related framework of the Quantum Eigenvalue Transformation (QET)~\cite{mizuta2024recursive,an2024laplace,dong2022ground}.

Alongside these algorithmic developments, there has been growing interest in the application of quantum computing to computer--aided engineering (CAE).  
CAE encompasses computational mechanics, chemistry, and physics, aiming to solve complex engineering problems through numerical simulations.  
However, classical CAE simulations often require enormous computational resources, particularly for large--scale systems involving nonlinear or multiphysics interactions.  
This computational demand motivates the exploration of quantum computing as a new paradigm for accelerating CAE applications.  
While computational chemistry and physics are expected to benefit greatly from quantum computers~\cite{zhang2025quantum,clinton2024towards,mcardle2020quantum,nielsen2010quantum}, recent studies have also demonstrated the feasibility of applying quantum algorithms to a broader class of computational mechanics problems formulated as partial differential equations (PDEs)~\cite{liu2024towards,jin2024quantum,raisuddin2024review}.  
For example, Hamiltonian simulation frameworks have demonstrated scalable quantum algorithms for solving PDEs arising from conserved physical systems~\cite{babbush2023exponential}.  
Related methodologies such as Linear Combinations of Hamiltonian Simulations (LCHS)~\cite{an2023linear} and Schr\"odingerization~\cite{jin2024quantum-in,jin2024quantum-sch,jin2023quantum} extend these approaches to a wider class of PDEs, including non--conserved systems.  
Furthermore, recent studies have highlighted the potential of quantum algorithms for nonlinear PDEs, leveraging Carleman linearization~\cite{krovi2023improved,liu2021efficient} and Koopman--von Neumann linearization~\cite{novikau2025quantum,higuchi2025quantum,joseph2020koopman}.  
Together, these developments shape the growing momentum toward Quantum CAE, where quantum resources are used to address the computational challenges of large--scale engineering simulations.

More specifically, in the field of computational mechanics within CAE, the Finite Element Method (FEM)~\cite{bonet2016nonlinear,belytschko2014nonlinear,hughes2012finite,de2011computational,onate2009structural} serves as a fundamental numerical framework for solving PDEs.
Quantum algorithms such as the Harrow--Hassidim--Lloyd (HHL) algorithm~\cite{duan2020survey,harrow2009quantum} have been proposed as quantum linear solvers, and Montanaro et al.~\cite{montanaro2016quantum} were among the first to apply HHL to FEM, marking an important milestone toward realizing Quantum CAE.  
However, FEM typically requires solving large-scale linear systems that yield symmetric positive definite (SPD) matrices, for which efficient and stable solvers are crucial.  
As the system size increases, these matrices often exhibit large condition numbers, significantly affecting numerical stability and computational efficiency.

Quantum linear system algorithms, including the HHL algorithm~\cite{harrow2009quantum} based on quantum phase estimation and the more general QSVT framework~\cite{martyn2021grand}, provide quantum methods for solving linear equations by effectively implementing matrix inversion.  
We collectively refer to these inverse--based approaches as the Quantum Matrix Inverse Algorithm (QMIA).  
Although QMIA offers a unified and elegant theoretical foundation, it faces a crucial practical limitation: the polynomial approximation of the inverse function $1/x$ becomes progressively more difficult as the condition number $\kappa$ increases.  
Near the lower end of the spectrum, $1/x$ exhibits steep growth, requiring QSP polynomials of degree $\mathcal{O}(\kappa \log(\kappa/\epsilon))$ to achieve a target 
accuracy~$\epsilon$~\cite{martyn2021grand}.
This large degree directly translates into deep quantum circuits, since each polynomial step requires calls to the block--encoding of $\bm{K}$ and controlled phase operations. 
As a result, for FEM matrices whose condition number grows with mesh refinement or material contrast, QMIA becomes increasingly resource--intensive and often impractical for quantum devices.

When we recall classical algorithms for solving linear systems, SPD systems are often solved using iterative approaches such as the steepest descent and conjugate gradient methods~\cite{scieur2016regularized,meza2010steepest,saad2003iterative,benzi2002preconditioning,eiermann2001geometric}, 
which are grounded in the variational minimization of quadratic energy functionals.  
These gradient--based methods are valued for their physical interpretability and numerical robustness, particularly in large--scale or ill--conditioned problems where direct inversion is impractical.  
Iterative solvers and preconditioning techniques of this kind have long been central to achieving efficient convergence in such systems.  

Analogous concepts are now being explored in the quantum domain.  
Raisuddin et al.~\cite{raisuddin2024qrls,raisuddin2024quantum} introduced quantum preconditioning strategies for quantum linear system algorithms, while Jin et al.~\cite{jin2024quantum-iter} developed a quantum iterative solver based on Schr\"odingerization.
More recently, Toyoizumi et al.~\cite{toyoizumi2024quantum} proposed a Quantum Conjugate Gradient method, further demonstrating the potential of quantum iterative approaches for SPD problems.
These advances collectively indicate a growing convergence between classical numerical methods and quantum algorithm design, suggesting that the next generation of quantum solvers will increasingly embrace variational and dynamical formulations rather than explicit matrix inversion.

Motivated by this convergence, we propose the Quantum Gradient Flow Algorithm (QGFA) as a novel quantum SPD linear system algorithms.
Unlike the QMIA, which explicitly performs matrix inversion through QSVT or phase estimation, QGFA is formulated on the basis of a variational principle and time--evolution dynamics.
By modeling the gradient--flow process of the corresponding quadratic energy functional, QGFA enables efficient convergence toward the solution without direct inversion, providing a promising alternative framework for quantum linear solvers that is both physically interpretable and less sensitive to the condition number.

The remainder of this paper is organized as follows.  
In Sec.~\ref{sec:method}, we present the formulation of the proposed QGFA.  
Section~\ref{subsec:SPD} introduces the variational principle of SPD linear systems and establishes the equivalence between the minimization problem and the corresponding linear system.  
QSP approximations of matrix functions for QGFA are introduced in Sec.~\ref{subsec:qsp}.
In Sec.~\ref{subsec:qet-lcu}, we describe the quantum circuit implementation of the gradient flow method using QET and the Linear Combination of Unitaries (LCU)~\cite{berry2015simulating,childs2012hamil,gribling2024optimal}.  
Section~\ref{sec:discussions} applies QGFA to FEM for solid mechanics problems and discusses the numerical results through tensile and cantilever beam simulations.  
Finally, Sec.~\ref{sec:conclusion} summarizes the findings and outlines future perspectives toward Quantum CAE.

The appendices provide supplementary materials supporting the proposed formulation:  
Appendix~\ref{sec:appendix-saf} introduces the soft absolute function, which enables even--function regularization of the target functions for QSP--based approximation in the QGFA time--evolution dynamics;  
Appendix~\ref{sec:appendix-fem} details the FEM formulation used for linear elastic deformation analysis;  
and Appendix~\ref{sec:appendix-inv} briefly reviews the QMIA for comparison with QGFA.

\section{Computational Method}\label{sec:method}

\subsection{SPD Linear Solvers as Time Evolution Systems}\label{subsec:SPD}

In this section, we introduce a variational principle for the SPD linear system and the characterization of the solution vector $\bm{u}$ as the minimizer of a variational energy functional. 
Let $\bm{K}\in\mathbb{R}^{n\times n}$ be a SPD matrix and $\bm{f}\in\mathbb{R}^n$ be a given vector. 
The corresponding variational principle is formulated by introducing the quadratic energy functional:
\begin{align}
    \Pi\left( \bm{u} \right) 
    = \frac{1}{2}\bm{u}^{T} \bm{K} \bm{u} - \bm{u}^{T}\bm{f}\ ,
    \label{eq:variational}
\end{align}
whose gradient and Hessian are:
\begin{align}
    \frac{\partial \Pi \left( \bm{u} \right)}{\partial \bm{u}} &= \bm{K}\bm{u} - \bm{f}\ ,
    \label{eq:pigrad}\\
    \frac{\partial^2 \Pi \left( \bm{u} \right)}{{\partial \bm{u}}^2} &= \bm{K}\ .
    \label{eq:pigradgrad}
\end{align}
Since the SPD matrix $\bm{K}$ has strictly positive eigenvalues, the functional $\Pi(\bm{u})$ is strictly convex and therefore admits a unique global minimizer.
In the context of FEM, $\bm{K}$ and $\bm{f}$ typically represent the stiffness matrix and the load vector, respectively.
The minimizer is characterized as follows:
\begin{align}
    \bm{u} = \arg\min_{\bm{v}}\Pi\left( \bm{v} \right) 
    &\;\;\Leftrightarrow\;\; 
    \frac{\partial \Pi \left( \bm{u} \right)}{\partial \bm{u}} = \bm{0}\nonumber\\
    &\;\;\Leftrightarrow\;\;
    \bm{K}\bm{u} = \bm{f}.
    \label{eq:argmin_u}
\end{align}
Equation~\eqref{eq:argmin_u} shows that solving the linear system $\bm{K}\bm{u}=\bm{f}$ is equivalent to minimizing the quadratic functional $\Pi\left( \bm{u} \right)$.
This equivalence provides the foundation for interpreting linear solvers in terms of variational principles of SPD systems.

In the algorithm of the steepest descent method~\cite{meza2010steepest}, the solution vector $\bm{u}$ is iteratively updated using the aforementioned facts.
However, we may also consider the time evolution of $\bm{u}(t)$, which moves forward toward the minimum of $\Pi\left( \bm{u} \right)$ using Eq.~\eqref{eq:pigrad}: 
\begin{align}
    \frac{d \bm{u}\left( t \right)}{d t} = - \frac{\partial \Pi \left( \bm{u} \right)}{\partial \bm{u}}= -\bm{K} \bm{u} + \bm{f}\ .
    \label{eq:time-evo-pigrad}
\end{align}
The analytical solution of Eq.~\eqref{eq:time-evo-pigrad} is given by:
\begin{align}
    \bm{u}\left( t \right) = e^{-\bm{K}t} \bm{u}\left( 0 \right) + \bm{K}^{-1}\left(\bm{I}-e^{-\bm{K}t}\right) \bm{f}\ ,
    \label{eq:gradient-flow-classical}
\end{align}
where $\bm{u}(0)$ is an initial solution vector.
Controlled by $t$, we regard $\bm{u}(t)$ in Eq.~\eqref{eq:gradient-flow-classical} as gradient--flow solution.
Taking the limit as $t\rightarrow \infty$, we obtain:
\begin{align}
    \bm{u}\left( t \right)|_{t\rightarrow \infty} = \bm{u}^{*} = \bm{K}^{-1} \bm{f}\ .
    \label{eq:invref}
\end{align}
With an appropriate choice of the initial vector $\bm{u}(0)$, Eq.~\eqref{eq:gradient-flow-classical} converges to the solution in Eq.~\eqref{eq:invref} within a relatively small time evolution $t$. 
The QET framework can be analyzed in terms of functions acting on each eigenvalue of $\bm{K}$.
Let us consider the eigenvalue decomposition as follows:
\begin{align}
    \bm{K} = \sum_i \lambda_i \ket{v_i}\bra{v_i}\,,
\end{align}
where $\lambda_i$ and $\ket{v_i}$ denote the $i$~th eigenvalue and corresponding eigenvector, respectively.  
To clarify the convergence behavior of the gradient--flow dynamics, we expand $\bm{u}(0)$, $\bm{f}$, $\bm{u}(t)$, and the exact solution $\bm{u}^*$ in the eigenbasis:
\begin{align}
    \bm{u}(0) &= \sum_i u_i^{v}\ket{v_i}\,,\\
    \bm{f} &= \sum_i f_i^{v}\ket{v_i}\,,\\
    \bm{u}(t) &= \sum_i \left( e^{-\lambda_i t} u_i^{v}
        + \frac{1 - e^{-\lambda_i t}}{\lambda_i} f_i^{v} \right)\ket{v_i}\,,\label{eq:classical-gradient-flow-eig-v}\\
    \bm{u}^{*} &= \sum_i \frac{f_i^{v}}{\lambda_i}\ket{v_i}\label{eq:classical-inv-eig-v}\,.
\end{align}
By appropriately rescaling $\bm{K}$, its eigenvalues can be mapped into the interval $\lambda_i \in [1/\kappa,\,1]$, where $\kappa$ denotes the condition number.
The error vector $\bm{\delta}(t) = \bm{u}(t) - \bm{u}^{*}$ then becomes as follows:
\begin{align}
    \bm{\delta}(t)
    &= \sum_i e^{-\lambda_i t}\left( u_i^{v} - \frac{f_i^{v}}{\lambda_i} \right)\ket{v_i}\,.
\end{align}
By taking the Euclidean norm and using $\lambda_i \ge 1/\kappa$, we can derive the error behavior as follows:
\begin{align}
    \|\bm{\delta}(t)\|
        &= \left( \sum_i e^{-2\lambda_i t} 
            \left| u_i^{v} - \frac{f_i^{v}}{\lambda_i} \right|^2
           \right)^{1/2} \nonumber\\
        &\le e^{-t/\kappa}
           \left( \sum_i 
            \left| u_i^{v} - \frac{f_i^{v}}{\lambda_i} \right|^2
           \right)^{1/2} \nonumber\\
        &= e^{-t/\kappa}\,\|\bm{\delta}(0)\|\,.
        \label{eq:delta-classic}
\end{align}
Eq.~\eqref{eq:delta-classic} shows that the gradient--flow solution $\bm{u}(t)$ converges exponentially to $\bm{u}^{*}$ at a rate governed by the smallest eigenvalue, resulting in a decay rate of $e^{-1/\kappa}$. 
To achieve an accuracy threshold $\zeta\geq\|\bm{\delta}(t)\|$, Eq.~\eqref{eq:delta-classic} implies the following sufficient condition for the time--evolution parameter:
\begin{align}
    t \ge -\kappa \ln\!\left( \frac{\zeta}{\|\bm{\delta}(0)\|} \right)\,.
    \label{eq:t-kappa-zeta-condition}
\end{align}
This shows that if the initial vector $\bm{u}(0)$ is already close to the exact solution $\bm{u}^{*}$, then $\|\bm{\delta}(0)\|$ becomes small, and consequently the required time $t$ decreases.
As in Eqs.~\eqref{eq:gradient-flow-classical}, \eqref{eq:invref}, \eqref{eq:classical-gradient-flow-eig-v}, and \eqref{eq:classical-inv-eig-v}, a smaller value of $t$ keeps the target functions in QSP farther away from the inverse--type behavior and makes them smoother over the eigenbasis.
As a result, the QSP procedure (see Sec.~\ref{subsec:qsp}) only needs to approximate well--behaved exponential--type functions, rather than functions with strong singularity near $x=0$, which substantially simplifies the polynomial approximation.

\begin{table*}[t]
    \renewcommand{\arraystretch}{1.5}
    \centering
    \caption{Comparison between the quantum matrix inverse algorithm (QMIA) and the quantum gradient flow algorithm (QGFA). The target functions are converted to the phase factors using quantum signal processing (QSP).}
    \label{tab:methods}
    \begin{tabular}{c|c|c}
        & QMIA & QGFA \\
        \hline
        Input data 
        & $\bm{K},\, \bm{f}$ 
        & $\bm{K},\, \bm{f},\, \bm{u}(0)$ \\
        \hline
        Output data 
        & \ $\bm{u}^{*} = \bm{K}^{-1} \bm{f}$ \ 
        & \ $\bm{u}(t) = e^{-\bm{K}t}\bm{u}(0) + \bm{K}^{-1}\!\!\left(\bm{I} - e^{-\bm{K}t}\right)\!\!\bm{f}$ \ \\
        \hline
        QSP target function 
        & $1/x$ 
        & $e^{-xt},\ \left(1-e^{-xt}\right)/x$ \\
        \hline
    \end{tabular}
\end{table*}

\begin{figure*}[t]
    \centering
    \includegraphics[width=0.95\linewidth]{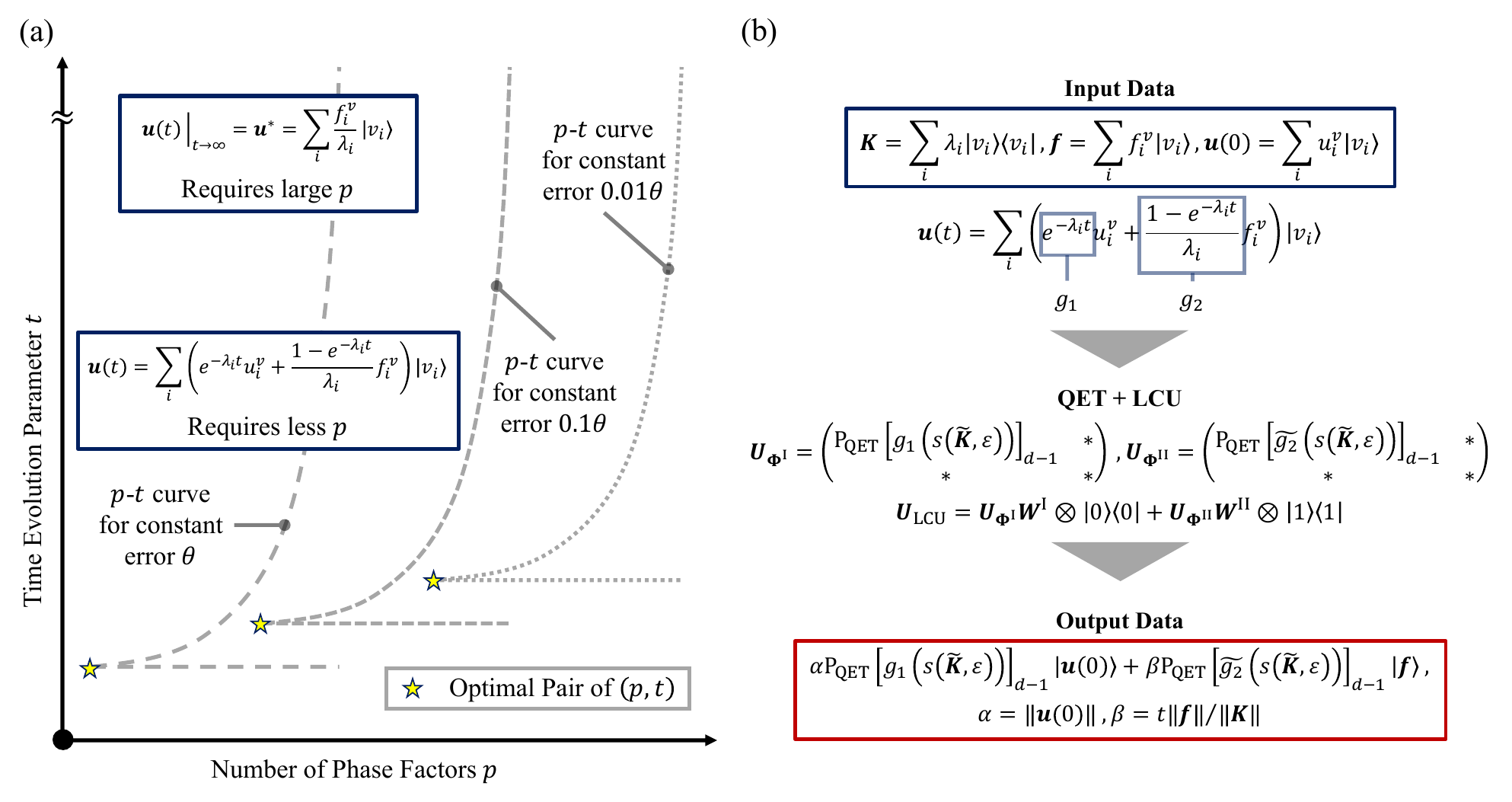}
    \caption{Schematic overview of the QGFA.
        (a) Relationship between the number of phase factors $p$ and the time--evolution parameter $t$ in the QGFA solution.
        For a fixed tolerance $\theta$, QGFA exhibits a characteristic $p$--$t$ curve.
        Along this curve, one can choose an optimal pair $(p,t)$ that attains the same accuracy as QMIA while requiring a significantly smaller number of phase factors $p$.
        (b) Flowchart of QGFA.
        The input consists of the SPD matrix $\bm{K}$, the load vector $\bm{f}$, and the initial vector $\bm{u}(0)$.
        Using the gradient--flow formulation, two matrix functions $g_{1}\!\left(s(\bm{K},\varepsilon)\right)$ and $\tilde{g}_{2}\!\left(s(\bm{K},\varepsilon)\right)$ are implemented through Quantum Eigenvalue Transformation (QET).  
        The Linear Combination of Unitaries (LCU) framework then coherently combines the two QET blocks,  
        producing the quantum state proportional to  
        $\alpha\,\text{P}\left[g_{1}(s(\bm{K},\varepsilon))\right]_{d-1}\,\bm{u}(0) + \beta\,\text{P}\left[\tilde{g}_{2}(s(\bm{K},\varepsilon))\right]_{d-1}\,\bm{f}$,  
        which corresponds to the gradient--flow solution. 
    }
    \label{fig:QGF-flow}
\end{figure*}

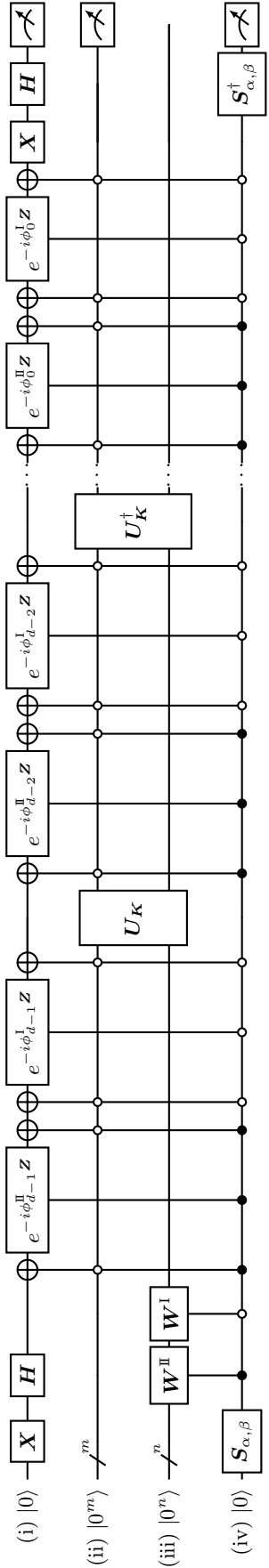
\begin{sidewaysfigure}
    \centering
    \begin{quantikz}[row sep=0.4cm, column sep=0.095cm]
        %--------------------------------------------------
        % QSP / parity ancilla  (H_QSP acts here)
        %--------------------------------------------------
        \lstick{(i) $\ket{0}$}
            & \gate{\bm{X}}
            & \gate{\bm{H}} % H_QSP: prepare |+> for parity (even/odd) interference
            & %\gate[wires=3]{\bm{U}_{\Phi^{\text{I\hspace{-1.2pt}I}}}}
            & \targ{}
            & \gate{e^{- i \phi^{\text{I\hspace{-1.2pt}I}}_{d-1}\bm{Z}}}%\gate[wires=2]{e^{\phi^{\text{I\hspace{-1.2pt}I}}_{d-1}(2\bm{\Pi}-\bm{I})}} % {R_{z}\left(\phi^{\mathrm{I\hspace{-1.2pt}I}}_{d-1}\right)}%\gate[wires=2]{e^{\phi^{\text{I\hspace{-1.2pt}I}}_{d-1}(2\bm{\Pi}-\bm{I})}}%\gate[wires=2]{\text{CR}(\phi_{d-2}^{\text{I\hspace{-1.2pt}I}})}
            & \targ{}
            & \targ{}
            & \gate{e^{- i \phi^{\text{I}}_{d-1}\bm{Z}}} %\gate[wires=2]{e^{\phi^{\text{I}}_{d-1}(2\bm{\Pi}-\bm{I})}}%{R_{z}\left(\phi^{\mathrm{I}}_{d-1}\right)}%\gate[wires=3]{\bm{U}_{\Phi^{\text{I}}}}
            & \targ{}
            & \qw
            & \targ{}
            & \gate{e^{- i \phi^{\text{I\hspace{-1.2pt}I}}_{d-2}\bm{Z}}}%\gate[wires=2]{e^{\phi^{\text{I\hspace{-1.2pt}I}}_{d-2}(2\bm{\Pi}-\bm{I})}} %{R_{z}\left(\phi^{\mathrm{I\hspace{-1.2pt}I}}_{d-2}\right)}
            & \targ{}
            & \targ{}
            & \gate{e^{- i \phi^{\text{I}}_{d-2}\bm{Z}}} %\gate[wires=2]{e^{\phi^{\text{I}}_{d-2}(2\bm{\Pi}-\bm{I})}}%{R_{z}\left(\phi^{\mathrm{I}}_{d-2}\right)}
            & \targ{}
            & 
            & \cdots
            & \targ{}
            & \gate{e^{- i \phi^{\text{I\hspace{-1.2pt}I}}_{0}\bm{Z}}} %\gate[wires=2]{e^{\phi^{\text{I\hspace{-1.2pt}I}}_{0}(2\bm{\Pi}-\bm{I})}}%{R_{z}\left(\phi^{\mathrm{I\hspace{-1.2pt}I}}_{0}\right)}
            & \targ{}
            & \targ{}
            & \gate{e^{- i \phi^{\text{I}}_{0}\bm{Z}}} %\gate[wires=2]{e^{\phi^{\text{I}}_{0}(2\bm{\Pi}-\bm{I})}}%{R_{z}\left(\phi^{\mathrm{I}}_{0}\right)}
            & \targ{}
            & \gate{\bm{X}}
            % & 
            % & \qw
            % & %\gate[wires=2]{e^{\phi^{\text{I\hspace{-1.2pt}I}}_{d-1}(2\bm{\Pi}-\bm{I})}}%\gate[wires=2]{\text{CR}(\phi_{0}^{\text{I\hspace{-1.2pt}I}})}
            & \gate{\bm{H}}
                % second H_QSP: measurement in {|+>, |->} basis
            & \meter{} \\
        %--------------------------------------------------
        % block-encoding ancilla + system register
        % (U_K acts here; QET for g1)
        %--------------------------------------------------
        \lstick{(ii) $\ket{0^{m}}$}
            & \qwbundle{m}
            & %\gate{e^{i\phi_{d-1}^{\text{I\hspace{-1.2pt}I}}}}
            & %\gate[wires=2]{{\bm{U}}_{\bm{K}}}
            & \ctrl[open]{-1}
            & %\gate{e^{i\phi_{d-2}^{\text{I\hspace{-1.2pt}I}}}}
            & \ctrl[open]{-1}
            & \ctrl[open]{-1}
            & %\gate[wires=2]{{\bm{U}}_{\bm{K}}}
            & \ctrl[open]{-1}
            & \gate[wires=2]{\bm{U}_{\bm{K}}}
            & \ctrl[open]{-1}
            &
            & \ctrl[open]{-1}
            & \ctrl[open]{-1}
            &
            & \ctrl[open]{-1}
            & \gate[wires=2]{\bm{U}_{\bm{K}}^{\dagger}}
            & \cdots
            & \ctrl[open]{-1}
            &
            & \ctrl[open]{-1}
            & \ctrl[open]{-1}
            &
            & \ctrl[open]{-1}
            &
            % & 
            % &
            % & %\gate{e^{i\phi_{0}^{\text{I\hspace{-1.2pt}I}}}}
            & \qw
            & \meter{}
            \\
        \lstick{(iii) $\ket{0^n}$}
            & \qwbundle{n}
            & \gate{\bm{W}^{\text{I\hspace{-1.2pt}I}}}
            & \gate{\bm{W}^{\text{I}}}
            &
            &
            &
            &
            &
            &
            &
            &
            &
            &
            &
            &
            &
            &
            & \cdots
            &
            &
            &
            &
            &
            &
            &
            % & 
            % &
            % &
            & \qw
            & \qw
            \\
        %--------------------------------------------------
        % LCU control qubit: selects g1- or g2-QET block
        %--------------------------------------------------
        \lstick{(iv) $\ket{0}$}
            & \gate{\bm{S}_{\alpha,\beta}}
            & \ctrl{-1}
            & \ctrl[open]{-1}
            & \ctrl{-2}
            & \ctrl{-3}
            & \ctrl{-2}
            & \ctrl[open]{-2}
            & \ctrl[open]{-3}
            & \ctrl[open]{-2}
            &
            & \ctrl{-2}
            & \ctrl{-3}
            & \ctrl{-2}
            & \ctrl[open]{-2}% \ctrl{-2}
            & \ctrl[open]{-3}
            & \ctrl[open]{-2}
            & % \ctrl{-2}
            & \cdots
            & \ctrl{-2}
            & \ctrl{-3}
            & \ctrl{-2}
            & \ctrl[open]{-2}
            & \ctrl[open]{-3}
            & \ctrl[open]{-2}
            &
            % &
            % & \qw
            % & \qw
            & \gate{\bm{S}_{\alpha,\beta}^{\dagger}}
            & \meter{}
    \end{quantikz}
   \caption{
        Quantum circuit for the QGFA. 
        From top to bottom, the four wires represent: 
        (i) the QSP ancilla qubit, where the Pauli--$X$ and Hadamard gates generate the even--odd interference needed to extract the imaginary part of the QET polynomial, 
        (ii) the block-encoding ancilla initialized in $\ket{0^{m}}$ and used to implement the block--encoding $\bm{U}_{\bm{K}}$ of $\tilde{\bm{K}}$, 
        (iii) the system register initialized in $\ket{0^{n}}$, on which the state-preparation unitaries $\bm{W}^{\mathrm{I}}$ and $\bm{W}^{\mathrm{II}}$ prepare $\ket{\bm{u}(0)}$ and $\ket{\bm{f}}$ and where the QET uniraty operators act, and 
        (iv) the LCU control qubit, which is rotated by $\bm{S}_{\alpha,\beta}$ to coherently select between the two QET sequences with phase sets $\bm{\Phi}^{\mathrm{I}}$ and $\bm{\Phi}^{\mathrm{II}}$. 
    }
    \label{fig:qc-for-qgfa}
\end{sidewaysfigure}

\subsection{Matrix Function Approximation using QSP}\label{subsec:qsp}

As shown in Table~\ref{tab:methods}, previous studies on QLSA~\cite{martyn2021grand,gilyen2019quantum,gribling2024optimal,raisuddin2024qrls,raisuddin2024quantum} have utilized the quantum matrix inverse algorithm (QMIA), which approximates the function $1/x$ using QSP.
However, the QMIA requires a large number of phase factors in the QSP due to its dependence on the condition number. 
Based on Sec.~\ref{subsec:SPD}, we instead consider a quantum gradient flow algorithm that approximates $e^{-xt}$ and $\left(1-e^{-xt}\right)/x$, with the aim of alleviating the dependence on the condition number.
The QGFA can be extended to a quantum algorithm using QET and LCU, as illustrated in Fig.~\ref{fig:QGF-flow} and \ref{fig:qc-for-qgfa}. 
% We propose this quantum version as the \textit{Quantum Gradient Flow Algorithm} (QGFA). 
To obtain Eq.~\eqref{eq:classical-gradient-flow-eig-v} in a quantum setting, the following functions $g_1(x)$ and $g_2(x)$ are approximated using QSP:
\begin{align}
    g_1(x) = e^{-xt}\,, \,g_2(x) = \frac{1-e^{-xt}}{x}\,.
    \label{eq:gfa-func}
\end{align}

To implement these functions via QSP, we take into account two fundamental restrictions as follows:

\begin{enumerate}
    \item The target function $f(x)$ must satisfy the bound $|f(x)|\leq 1$ for all $x\in[-1,1]$.
    \item The target function must be either an even or odd function, 
    and sufficiently smooth so that it admits an accurate polynomial approximation 
    (e.g., Chebyshev approximation) over the domain of interest.
\end{enumerate}

The first restriction is satisfied for $g_1(x)$ but not for $g_2(x)$. 
Since an SPD system has only positive eigenvalues, we consider only the region $x > 0$. 
In this case, $g_2(x)$ attains its maximum value at $x=0$. 
Taking the limit as $x \rightarrow 0$, we obtain:
\begin{align}
    \lim_{x\rightarrow 0} g_2(x) = &\lim_{x\rightarrow 0} \left\{ t - \frac{t^2 x}{2!} + \frac{t^3 x^2}{3!}- \cdots \right. \nonumber
    \\ &\left.+(-1)^{n-1}\frac{t^{n} x^{n-1}}{n!}+ \cdots \right\} = t
\end{align}
Here, the time evolution parameter $t$ takes a positive value. 
To satisfy the first restriction, we normalize $g_2(x)$ by introducing $\tilde{g_2}(x)$ for QSP as follows:
\begin{align}
    \tilde{g_2}(x)=\frac{g_2(x)}{t}\,.
\end{align}
To satisfy the second restriction, we employ the soft absolute function $s(x,\varepsilon)$, which symmetrizes the argument and yields an even, $C^{\infty}$--smooth function:
\begin{align}
    &s(x,\varepsilon) = \varepsilon \left\{ \ln \left( e^{\frac{x}{\varepsilon}} +e^{-\frac{x}{\varepsilon}} \right) \right\}\,,\\
    &\lim_{\varepsilon\rightarrow 0} s(x,\varepsilon) = |x|\ (|x|\gg \varepsilon),
\end{align}
where $\varepsilon$ is a softening parameter.
The detailed properties of the soft absolute function are discussed in Appendix~\ref{sec:appendix-saf}.
We obtain the appropriate functions ${g_1}(s(x,\varepsilon))$ and $\tilde{g_2}(s(x,\varepsilon))$ for QSP as follows:
\begin{align}
    &{g_1}(s(x,\varepsilon)) = e^{-s(x,\varepsilon)t}\,, \label{eq:target-qsp-final}\\
    &\tilde{g_2}(s(x,\varepsilon)) = \frac{g_2(s(x,\varepsilon))}{t}=\frac{1-e^{-s(x,\varepsilon)t}}{s(x,\varepsilon)t}\,.
    \label{eq:target-qsp-final-g2}
\end{align}
To determine an appropriate value of the smoothing parameter $\varepsilon$, we formulate a nonlinear optimization problem 
based on the relative error between the soft absolute function $s(x,\varepsilon)$ and the exact absolute value $|x|$. 
The relative errors of the target functions ${g_1}$ and $\tilde{g_2}$ are defined as
\begin{align}
    r_1(x,\varepsilon)&=\frac{|{g_1}(s(x,\varepsilon))-g_1(|x|)|}{g_1(|x|)}\nonumber\\
    &=1-e^{-\left\{s(x,\varepsilon) - |x| \right\}t}\leq\eta\,,\\
    r_2(x,\varepsilon)&=\frac{|\tilde{g_2}(s(x,\varepsilon))-\tilde{g_2}(|x|)|}{\tilde{g_2}(|x|)}\nonumber\\
    &=1-\frac{1-e^{-s(x,\varepsilon)t}}{1-e^{-|x|t}}\left\{\frac{s(x,\varepsilon)}{|x|}\right\}^{-1}\leq\eta\,,
\end{align}
where $\eta$ denotes the prescribed upper bound of the relative error.
From Eq.~\eqref{eq:abserror-softabs}, it follows that both $r_1(x,\varepsilon)$ and $r_2(x,\varepsilon)$ are monotonically decreasing with respect to $x$.
Therefore, the optimal smoothing parameter $\varepsilon$ can be determined by solving the following nonlinear equations:
\begin{align}
    r_i\left(\frac{1}{\kappa},\varepsilon\right) = \eta\, \,(i=1,2)\,, \label{eq:solve-eps}
\end{align}
where $\kappa$ is the condition number of the aforementioned SPD system.

\subsection{Quantum Circuit Representation of QGFA using QET and LCU}\label{subsec:qet-lcu}

By using the QSP introduced in Sec.~\ref{subsec:qsp}, the full quantum circuit can be constructed using QET and LCU, as shown in Fig.~\ref{fig:qc-for-qgfa}.
We prepare the total Hilbert space $\mathcal{H}$ in the QET and LCU implementation as follows:
\begin{align}
    &\mathcal{H}=\mathcal{H}_{\mathrm{QSP}}
    \otimes\mathcal{H}_{\mathrm{BE}}
    \otimes\mathcal{H}_{\mathrm{sys}}\otimes\mathcal{H}_{\mathrm{LCU}}\,,\\
    &\mathcal{H}_{\mathrm{QSP}}\in\mathbb{C}^2\,,\,\mathcal{H}_{\mathrm{BE}}\in\left(\mathbb{C}^2\right)^{\otimes m}\,,\nonumber\\
    &\mathcal{H}_{\mathrm{sys}}\in\left(\mathbb{C}^2\right)^{\otimes n}\,,\mathcal{H}_{\mathrm{LCU}}\in\mathbb{C}^2\,.\nonumber
\end{align}
where $\mathcal{H}_{\mathrm{QSP}}$, $\mathcal{H}_{\mathrm{BE}}$, $\mathcal{H}_{\mathrm{sys}}$, and $\mathcal{H}_\mathrm{LCU}$ represent the single--qubit ancilla space used for the imaginary--part extraction in the QSP construction, 
$m$--qubit register used for the block--encoding, the Hilbert space on which $\bm{K}$ acts, and single--qubit ancilla space used for LCU, respectively.

To begin with, we consider the Hilbert space of $\mathcal{H}_{\mathrm{QSP}}\otimes\mathcal{H}_{\mathrm{BE}}\otimes\mathcal{H}_{\mathrm{sys}}$ for QET.
For the unitary block--encoding of $\bm{K}$, we assume primary access to $\bm{U}_{\bm{K}}$ acting on $\mathcal{H}_{\mathrm{BE}}\otimes\mathcal{H}_{\mathrm{sys}}$ as follows:
\begin{align}
  \bm{U}_{\bm{K}}
  =
  \begin{pmatrix}
    \tilde{\bm{K}} & *\\
    * & *
  \end{pmatrix}\,,\,
  \tilde{\bm{K}} = \frac{\bm{K}}{\|\bm{K}\|}\,.
\end{align}
We extend this to the space $\mathcal{H}_{\mathrm{QSP}}\otimes\mathcal{H}_{\mathrm{BE}}\otimes\mathcal{H}_{\mathrm{sys}}$ as follows:
\begin{align}
  \widetilde{\bm{U}}_{\bm{K}}
  =
  \bm{I}_{\mathrm{QSP}}\otimes\bm{U}_{\bm{K}}\,,
\end{align}
where $\bm{I}_{\mathrm{QSP}}$ is the identity operator in $\mathcal{H}_{\mathrm{QSP}}$.
In this study, we utilized \texttt{pyQSP}~\cite{martyn2021grand,gilyen2019quantum,dong2021efficient,chao2020finding,haah2019product} to obtain the phase factors 
$\bm{\Phi}^{\text{I}} = \left(\phi_0^{\text{I}},\phi_1^{\text{I}},\cdots, \phi_{d-1}^{\text{I}}\right)\in\mathbb{R}^{d}$ and 
$\bm{\Phi}^{\text{I\hspace{-1.2pt}I}} = \left(\phi_0^{\text{I\hspace{-1.2pt}I}},\phi_1^{\text{I\hspace{-1.2pt}I}},\cdots, \phi_{d-1}^{\text{I\hspace{-1.2pt}I}}\right)\in\mathbb{R}^{d}$ 
for the QSP of Eq.~\eqref{eq:target-qsp-final} and \eqref{eq:target-qsp-final-g2}.
The QET procedure yields the corresponding matrix functions in the form of unitary operators 
$\bm{U}_{\bm{\Phi}^{\text{I}}}$ and $\bm{U}_{\bm{\Phi}^{\text{I\hspace{-1.2pt}I}}}$, defined as follows:
\begin{align}
    \bm{U}_{\bm{\Phi}^{\text{I}}} = & \bm{R}_{z}\left(\phi^{{\text{I}}}_{0} \right)\prod_{j=1}^{(d-1)/2} \left[\widetilde{\bm{U}}^{\dagger}_{\bm{K}} \bm{R}_{z}\left(\phi^{{\text{I}}}_{2j-1} \right)\widetilde{\bm{U}}_{\bm{K}} \bm{R}_{z}\left(\phi^{{\text{I}}}_{2j} \right) \right]\nonumber\\
    = & \left(
    \begin{matrix}
        \text{P}_\text{QET}\left[g_{1}\left( s\left( \tilde{\bm{K}},\varepsilon \right) \right)\right]_{d-1} & * \\
        * & *
    \end{matrix}
    \right)\,,\label{eq:QET1}\\ 
    \bm{U}_{\bm{\Phi}^{\text{I\hspace{-1.2pt}I}}} =& \bm{R}_{z}\left(\phi^{\text{I\hspace{-1.2pt}I}}_{0}\right)\prod_{j=1}^{(d-1)/2} \left[\widetilde{\bm{U}}^{\dagger}_{\bm{K}} \bm{R}_{z}\left(\phi_{2j-1}^{\text{I\hspace{-1.2pt}I}}\right) \widetilde{\bm{U}}_{\bm{K}} \bm{R}_{z}\left(\phi_{2j}^{\text{I\hspace{-1.2pt}I}}\right) \right]\nonumber\\
    = & \left(
    \begin{matrix}
        \text{P}_\text{QET}\left[\tilde{g_2}\left( s\left( \tilde{\bm{K}},\varepsilon \right) \right)\right]_{d-1}  & * \\
        * & *
    \end{matrix}
    \right)\,,\label{eq:QET2}\\
    \bm{C}_{0^{m}} = & \bm{I}_{\mathrm{QSP}}\otimes \left( \bm{I}_{\mathrm{BE}} - \ket{0^m}\bra{0^m}\right)\otimes\bm{I}_{\text{sys}} \nonumber \\
    &+ \bm{X}\otimes \ket{0^m}\bra{0^m}\otimes\bm{I}_{\text{sys}}\,, \label{eq:controllnot-zero}\\
    \bm{R}_{z}\left( \phi \right)  = & \bm{C}_{0^m} \left(e^{-i \phi \bm{Z}}\otimes \bm{I}_{\mathrm{BE}}\otimes \bm{I}_{\mathrm{sys}}\right) \bm{C}_{0^m} \label{eq:controll-rz}\,,
\end{align}
where $\bm{I}_{\mathrm{BE}}$, $\bm{I}_\mathrm{sys}$, $\bm{X}$, $\bm{Z}$, $\bm{C}_{0^m}$, $\bm{R}_z\left(\phi\right)$, and $\mathrm{P}_{\mathrm{QET}}[\cdot]_{d-1}$ denote
the identity operator on $\mathcal{H}_{\mathrm{BE}}$, 
the identity operator on $\mathcal{H}_{\mathrm{sys}}$,
the Pauli--$X$ gate,
the Pauli--$Z$ gate,
controlled--on--$\ket{0}$ NOT gate on $\mathcal{H}_{\mathrm{QSP}}\otimes\mathcal{H}_{\mathrm{BE}}$,
controlled--on--$\ket{0}$ $Z$--rotation gate, 
and the $(d-1)$--degree QET polynomial acting on $\mathcal{H}_{\mathrm{sys}}$, respectively.
In this formulation, the top--left block of each unitary operator 
$\bm{U}_{\bm{\Phi}^{\text{I}}}$ and $\bm{U}_{\bm{\Phi}^{\text{II}}}$ acts on the 
subspace $\ket{0}\otimes\ket{0^m}\otimes\mathcal{H}_{\mathrm{sys}}$ and implements a 
QSP--based polynomial transformation of the normalized operator $\tilde{\bm{K}}$.
Consequently, these blocks approximate the target matrix functions 
$g_1\!\left(s(\tilde{\bm{K}},\varepsilon)\right)$ and 
$\tilde{g}_2\!\left(s(\tilde{\bm{K}},\varepsilon)\right)$ via their QET polynomial 
representations as summarized in Eqs.~\eqref{eq:QET1} and \eqref{eq:QET2}.
By applying Pauli--$X$ gate $\bm{X}$ and Hadamard gate $\bm{H}$ to the ancilla in $\mathcal{H}_{\mathrm{QSP}}$, 
the real polynomial approximation
% $\mathrm{P}[\cdot]$
is extracted from the 
imaginary part of $\mathrm{P}_{\mathrm{QET}}[\cdot]$~\cite{martyn2021grand,gilyen2019quantum}.  
Note that Eqs.~\eqref{eq:QET1} and \eqref{eq:QET2} correspond to the standard QET construction when $d$ is odd.  
The same polynomial transformations are obtained for even $d$ as well~\cite{martyn2021grand,gilyen2019quantum}.

To obtain the the equivalent results with Eqs.~\eqref{eq:gradient-flow-classical} and \eqref{eq:classical-gradient-flow-eig-v} in quantum algorithm, the summation of the terms are required as follows: 
\begin{align}
    \alpha &\text{P}_{\text{QET}}\left[g_{1}\left( s\left( \tilde{\bm{K}},\varepsilon \right) \right)\right]_{d-1}\ket{\bm{u}(0)}\nonumber
    \\&+
    \beta \text{P}_{\text{QET}}\left[\tilde{g_{2}}\left( s\left( \tilde{\bm{K}},\varepsilon \right) \right)\right]_{d-1}\ket{\bm{f}}\,,\label{eq:post-result}\\
    \alpha&=\|\bm{u}(0)\|\,,\,\beta=\frac{t\|\bm{f}\|}{\|\bm{K}\|}\,\nonumber.
\end{align}
Remind that $\bm{u}(0)$ and $\bm{f}$ are given in the classical formulation. 
These vectors can be straightforwardly extended to the corresponding quantum state vectors $\ket{\bm{u}(0)}$ and $\ket{\bm{f}}$, 
which are normalized such that $\langle \bm{u}(0) | \bm{u}(0) \rangle = \langle \bm{f} | \bm{f} \rangle = 1$, 
% and have the same dimension as the unitary operators $\bm{U}_{\bm{\Phi}^{\text{I}}}$ and $\bm{U}_{\bm{\Phi}^{\text{I\hspace{-1.2pt}I}}}$
respectively.
The summation in Eq.~\eqref{eq:post-result} can be realized by LCU using the ancilla qubits on $\mathcal{H}_{\mathrm{LCU}}$.

The single ancilla qubit $\mathcal{H}_{\mathrm{LCU}}$ plays a crucial role in coherently selecting between the two QET unitarys
$\bm{U}_{\bm{\Phi}^{\mathrm{I}}}$ and $\bm{U}_{\bm{\Phi}^{\text{I\hspace{-1.2pt}I}}}$.
Specifically, controlled operations in the circuit ensure that the phase sequence $\bm{\Phi}^{\mathrm{I}}$ is applied only when the LCU qubit is in the
state $\ket{0}$, whereas the sequence
$\bm{\Phi}^{\text{I\hspace{-1.2pt}I}}$ is applied only when the LCU qubit is in the
state $\ket{1}$.
Furthermore, prior to invoking the QET blocks, the system register is initialized to $\ket{0^{n}}$.  
We introduce two state-preparation unitaries $\bm{W}^{\mathrm{I}}$ and 
$\bm{W}^{\mathrm{II}}$ acting solely on the system Hilbert space 
$\mathcal{H}_{\mathrm{sys}}$, which prepare the required input states as follows:
\begin{align}
    \bm{W}^{\mathrm{I}}\ket{0^{n}} = \ket{\bm{u}_{0}}\,,\,
    \bm{W}^{\mathrm{II}}\ket{0^{n}} = \ket{\bm{f}}\,.
\end{align}
Including these state preparation unitaries,
we can prepare an overall block--diagonal unitary $\bm{U}_{\mathrm{LCU}}$ as follows:
\begin{align}
    \bm{U}_{\mathrm{LCU}}
    =  & \bm{U}_{\bm{\Phi}^{\text{I}}}\widetilde{\bm{W}}^{\mathrm{I}} \otimes \ket{0}\bra{0}
    + \bm{U}_{\bm{\Phi}^{\text{I\hspace{-1.2pt}I}}} \widetilde{\bm{W}}^{\mathrm{\text{I\hspace{-1.2pt}I}}} \otimes \ket{1}\bra{1}\,,\label{eq:LCU-block}\\
    \widetilde{\bm{W}}^{\mathrm{I}} = & \bm{I}_{\mathrm{QSP}}\otimes\bm{I}_{\mathrm{BE}}\otimes \bm{W}^{\mathrm{I}}\,,\nonumber\\
     \widetilde{\bm{W}}^{\mathrm{\text{I\hspace{-1.2pt}I}}} = & \bm{I}_{\mathrm{QSP}}\otimes\bm{I}_{\mathrm{BE}}\otimes  \bm{W}^{\mathrm{\text{I\hspace{-1.2pt}I}}}\,.\nonumber
\end{align}
Here, both unitaries $\bm{U}_{\bm{\Phi}^{\mathrm{I}}}$ and $\bm{U}_{\bm{\Phi}^{\mathrm{II}}}$ are applied conditionally on the LCU ancilla states $\ket{0}$ and $\ket{1}$, respectively.  
Taking a closer look at their internal structure, each QET unitary consists of alternating applications of the block--encoding $\bm{U}_{\bm{K}}$ and single--qubit phase rotations determined by the phase sequences $\bm{\Phi}^{\mathrm{I}}$ or $\bm{\Phi}^{\mathrm{II}}$.
In principle, $\bm{U}_{\bm{K}}$ must also be controlled on $\ket{0}$ and $\ket{1}$ in accordance with the LCU branching.
However, a key simplification arises: a controlled--on--$\ket{0}$ application of $\bm{U}_{\bm{K}}$ and a controlled--on--$\ket{1}$ application of $\bm{U}_{\bm{K}}$ together are equivalent to an unconditional application of $\bm{U}_{\bm{K}}$.  
That is, the two controlled versions multiply to the same action as the bare block--encoding.  
As a result, only the phase--rotation gates in the QSP sequence need to be controlled by the LCU ancilla, while $\bm{U}_{\bm{K}}$ itself can be applied without any control.
Consequently, as illustrated in Fig.~\ref{fig:qc-for-qgfa}, the phase sequences $\bm{\Phi}^{\mathrm{I}}\in\mathbb{R}^{d}$ and $\bm{\Phi}^{\mathrm{II}}\in\mathbb{R}^{d}$ act on $\bm{U}_{\bm{K}}$ in a controlled manner, but the block--encoding $\bm{U}_{\bm{K}}$ is invoked only $(d-1)$ times in each branch.

To coherently combine these two QET unitaries, the mixing unitary $\bm{S}_{\alpha,\beta}$ on $\mathcal{H}_{\mathrm{LCU}}$ is applied, defined as follows:
\begin{align}
    \bm{S}_{\alpha,\beta}
    &=
    \frac{1}{\sqrt{\alpha+\beta}}
    \begin{pmatrix}
        \sqrt{\alpha} & -\sqrt{\beta} \\
        \sqrt{\beta}  & \sqrt{\alpha}
    \end{pmatrix}\,.
\end{align}
In this way, measuring the LCU ancilla in the computational basis and post--selecting the outcome $\ket{0}$ yields the normalized state proportional to Eq.~\eqref{eq:post-result}.

As shown in Eqs.~\eqref{eq:QET1} and \eqref{eq:QET2}, each target function
is implemented using a QET polynomial of degree $(d-1)$.
For the overall calculation flow of QGFA in Fig.~\ref{fig:qc-for-qgfa},
two QET unitaries, $\bm{U}_{\bm{\Phi}^{\mathrm{I}}}$ and
$\bm{U}_{\bm{\Phi}^{\mathrm{II}}}$, are constructed, resulting in a
total of $2d$ phase factors in the QSP sequences.
However, the dominant computational cost in QET is not determined by the
number of phase factors, but rather by how many times the block--encoding
$\bm{U}_{\bm{K}}$ is applied.  
Although QGFA employs two separate QET blocks, each block invokes
$\bm{U}_{\bm{K}}$ (or $\bm{U}_{\bm{K}}^{\dagger}$) exactly $(d-1)$ times.
Because these calls are executed within a coherent LCU structure, the
number of block--encoding operations does not double.  
Consequently, the total block--encoding cost still scales as $d$ rather
than $2d$.

This is a practical advantage of QGFA: even though two QET matrix functions
are implemented and combined through LCU, the algorithm incurs
essentially the same block--encoding complexity as a single QET matrix function, while simultaneously reducing the dependence on the
condition number compared with QMIA.
% }

\section{Results \& Discussions}\label{sec:discussions}

\begin{figure*}[t]
    \centering
    \includegraphics[width=0.85\linewidth]{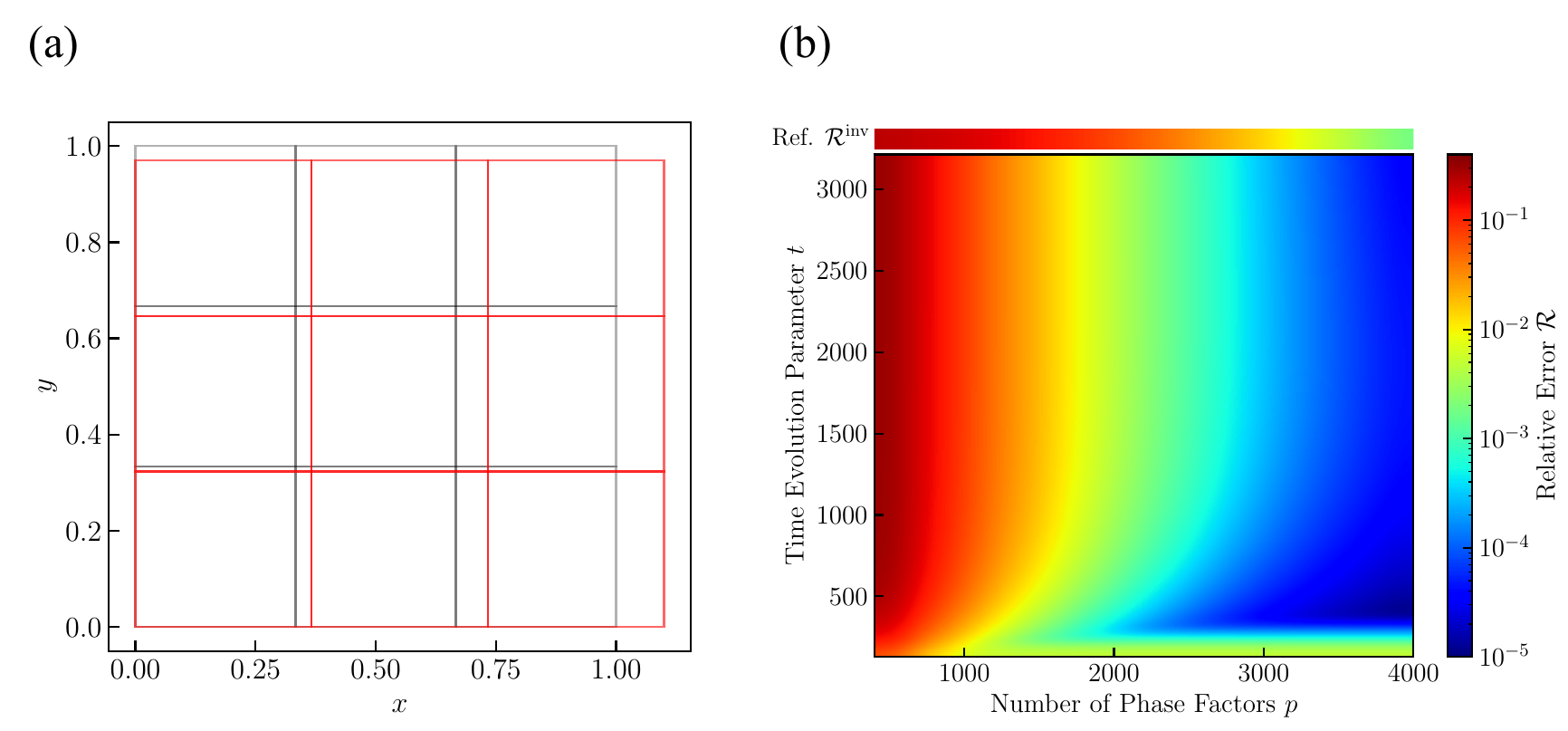}
    \caption{
    Tensile simulation with a $3\times3$ (9-element) quadrilateral mesh solved by the finite element method (FEM) using the QGFA. 
    (a) Geometry and discretization of the unit square ($x,y\in[0,1]$) with the tensile boundary condition; the remaining edges are traction-free. 
    (b) Distribution of the relative error $\mathcal{R}$ between the QGFA solution and the classical FEM solution, obtained by varying the time-evolution parameter $t$ (vertical axis) and the number of phase factors $p$ (horizontal axis). 
    For comparison, the reference relative error $\mathcal{R}^{\text{inv}}$ from the QMIA is also indicated.
    }
    \label{fig:results-9-ele-ten}
\end{figure*}

\begin{figure*}[t]
    \centering
    \includegraphics[width=0.85\linewidth]{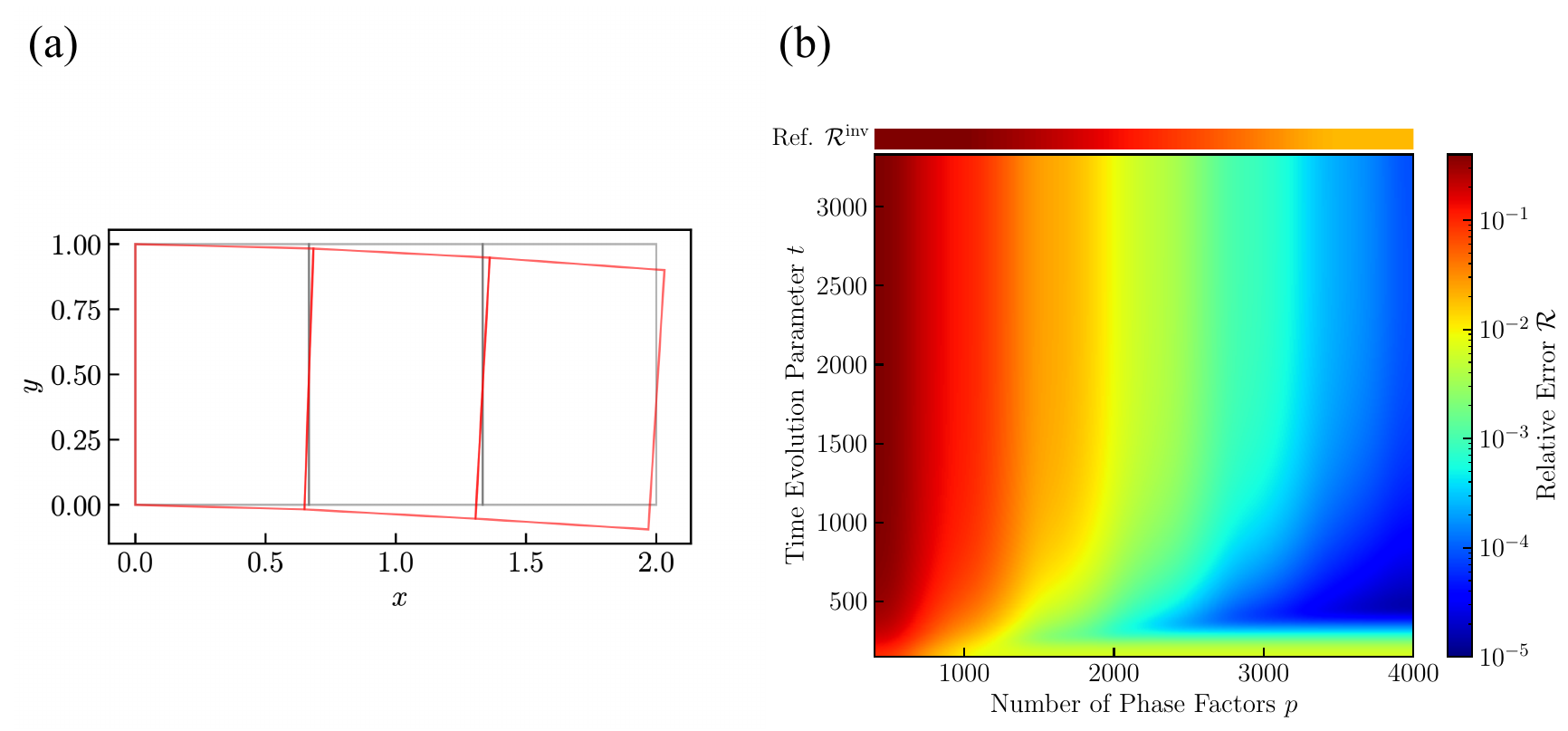}
    \caption{
    Cantilever beam bending simulation with 3 quadrilateral elements solved by the FEM using QGFA.
    (a) Geometry and boundary conditions of the cantilever beam: the left edge is fully fixed, and a vertical load is applied at the right edge. 
    (b) Distribution of the relative error $\mathcal{R}$ between the QGFA solution and the classical FEM solution, obtained by varying the time-evolution parameter $t$ (vertical axis) and the number of phase factors $p$ (horizontal axis).
    For comparison, the reference relative error $\mathcal{R}^{\text{inv}}$ obtained from the QMIA is also indicated.
    }
    \label{fig:results-3-ele-cant}
\end{figure*}

\begin{figure*}[t]
    \centering
    \includegraphics[width=0.85\linewidth]{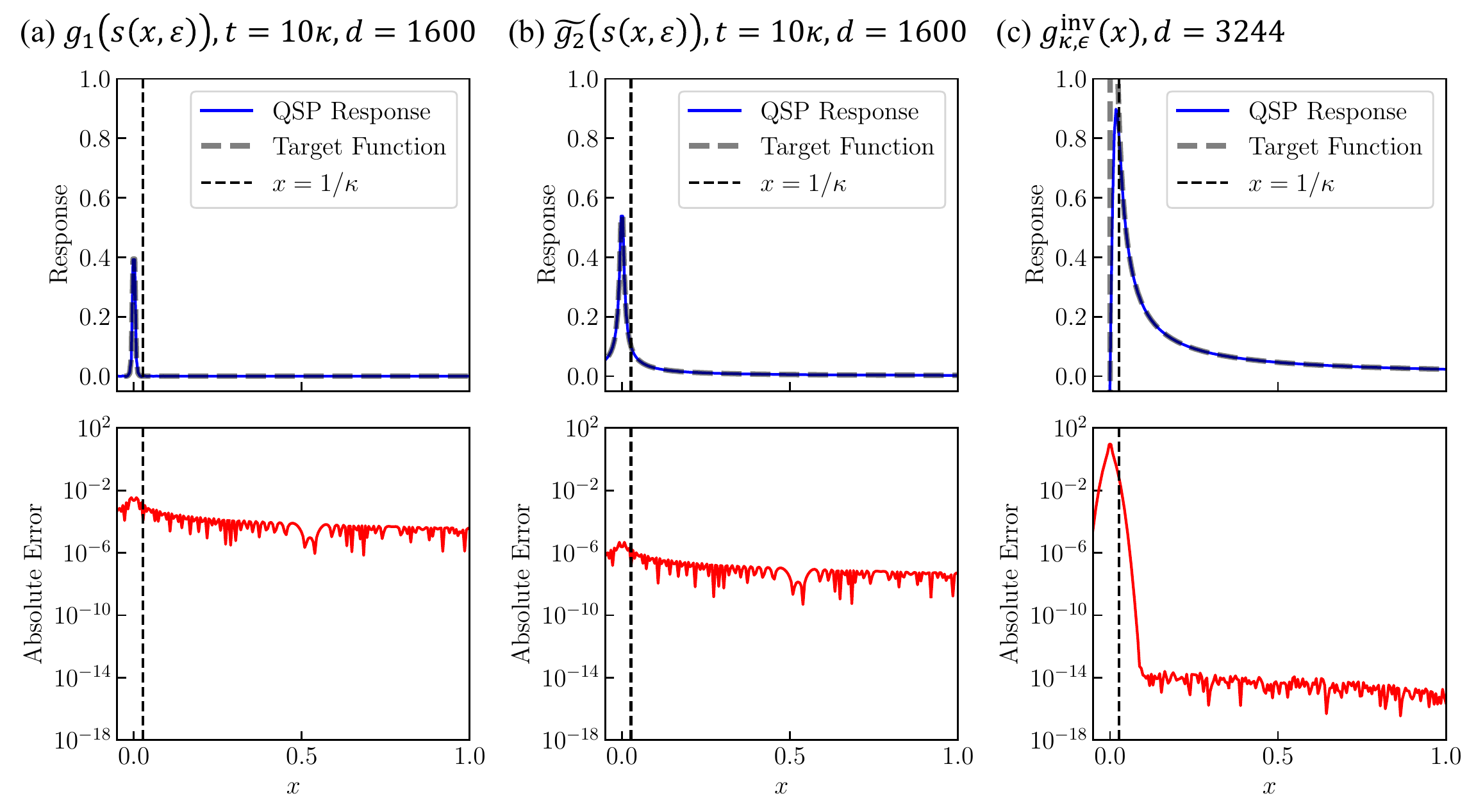}
    \caption{
    Results of the QSP response in cantilever beam bending simulation with 3 quadrilateral elements ($\kappa = 37.018$). Response of (a) $g_1\left(s(x,\varepsilon)\right)$, (b) $\tilde{g_2}\left(s(x,\varepsilon)\right)$, and (c) $g_{\kappa,\epsilon}^{\text{inv}}(x)$.
    }
    \label{fig:results-qsp-func}
\end{figure*}

In this section, we demonstrate the performance of the proposed QGFA through FEM problems.
The objective is to verify the numerical accuracy and convergence behavior 
of QGFA compared with QMIA.
% In this section, we present numerical examples of the FEM implemented using the proposed QGFA.
For simplicity, we focus on FEM deformation analyses in the field of solid mechanics.
Specifically, we consider two-dimensional plane stress problems of linear elastic materials, discretized using 4-node quadrilateral isoparametric elements~\cite{bonet2016nonlinear,belytschko2014nonlinear,hughes2012finite,de2011computational,onate2009structural}.
The detailed formulation of the stiffness equation is described in Appendix~\ref{sec:appendix-fem}.
We set Young’s modulus and Poisson’s ratio to $E = 0.2$ and $\nu = 0.3$, respectively, and perform all analyses on the $xy$-plane.

For the QGFA setup, the time-evolution parameter $t$ is determined based on Eq.~\eqref{eq:t-kappa-zeta-condition}, 
and the smoothing parameter $\varepsilon$ is obtained by solving Eq.~\eqref{eq:solve-eps}.
In these case studies, $\varepsilon$ is selected such that $\eta = 10^{-6}$.
As described in Eq.~\eqref{eq:QET1} and Eq.~\eqref{eq:QET2}, we prepare $(d-1)$ phase factors for each function.
The total number of phase factors $p$ required to construct QGFA is therefore $p = 2d$.

To verify the accuracy of our algorithm, we compute the relative error $\mathcal{R}$ between the QGFA solution $\bm{u}_{\text{qc}}(t)$ and the exact solution $\bm{u}^{*}$ as follows:
\begin{align}
    \mathcal{R} = \frac{||\bm{u}_{\text{qc}}(t) - \bm{u}^{*}||}{||\bm{u}^{*}||}\,.
\end{align}
In this simulation, we change the time evolution parameter $t$ and the number of phase factors $p$. 
% Eq.~\eqref{eq:QET1} and \eqref{eq:QET2} follows that $p=2d$ in QFGA.

For reference, we also solve the FEM using QMIA implemented with \texttt{pyQSP}~\cite{martyn2021grand,gilyen2019quantum,dong2021efficient,chao2020finding,haah2019product}, and compute the corresponding relative error:
\begin{align}
    \mathcal{R}^{\text{inv}} = \frac{||\bm{u}^{\text{inv}}_{\text{qc}} - \bm{u}^{*}||}{||\bm{u}^{*}||}\,,
\end{align}
where $\bm{u}_{\text{qc}}^{\text{inv}}$ is the displacement vector obtained from QMIA.
The detailed formulation of QMIA is described in Appendix~\ref{sec:appendix-inv}.
We change the number of phase factors $p$ and observe the difference compared to QGFA.

\subsection{Numerical Example: Tensile Simulation}

Figure~\ref{fig:results-9-ele-ten}(a) illustrates the setup and results 
for the 9-element tensile simulation.
We consider a square domain of $1.0\times1.0$, consisting of 16 nodes and 9 quadrilateral elements.
Displacement constraints are applied along the boundaries at $x=0$ and $y=0$, 
while a prescribed displacement of $0.1$ is applied at $x=1.0$ in the $x$-direction.
Under these conditions, the condition number of the stiffness matrix $\bm{K}$ is $\kappa = 32.136$.

Figure~\ref{fig:results-9-ele-ten}(b) shows the distribution of the relative error $\mathcal{R}$ obtained by varying $t$ and $p$. 
We provide a well--prepared initial vector $\bm{u}(0)$ that is close to the analytical solution $\bm{u}^{*}$ as explained in Appendix~\ref{sec:appendix-fem}. 
In the FEM formulation, the imposed displacement is incorporated into $\bm{u}(0)$ as described in Eq.~\eqref{eq:fem-input-qgfa}.
This initialization allows rapid convergence, achieving $\mathcal{R} = 10^{-4}$ even for a small time-evolution parameter $t \approx 500$ 
and a relatively small number of phase factors $p = 3000$.
For smaller values of $t$ (i.e., $t \leq 300$), the quantum state $\bm{u}_{\text{qc}}(t)$ has not yet evolved sufficiently toward $\bm{u}^{*}$.
In contrast, for larger $t$ values ($1000 \leq t \leq 3000$), the solution requires a higher number of phase factors $p$ compared with $t = 500$.
These results indicate that a small $t$ can lead to faster convergence when an appropriate initial state $\bm{u}(0)$ is provided.

The top color bar in Fig.~\ref{fig:results-9-ele-ten}(b) indicates the reference relative error $\mathcal{R}^{\text{inv}}$ obtained using QMIA.
Even when the number of phase factors is increased up to $p=4000$, the error of the QMIA remains at approximately $\mathcal{R}^{\mathrm{inv}}\!\approx 10^{-3}$.
This comparison suggests that when a suitable initial (hot-start) solution vector $\bm{u}(0)$ is used, 
QGFA can achieve accurate convergence with a smaller number of phase factors $p$.

\subsection{Numerical Example: Cantilever Beam Bending Simulation}

Figure~\ref{fig:results-3-ele-cant}(a) illustrates the setup and results 
for the 3-element cantilever beam problem consisting of 8 nodes.
We consider a rectangular domain of $1.0\times2.0$.
The left edge ($y=0$) is fully fixed, and a vertical displacement of $0.1$ is applied at $x=2.0$ in the $y$-direction.
The entire domain is discretized using 3 quadrilateral elements.
Under these conditions, the condition number of the stiffness matrix $\bm{K}$ is $\kappa = 37.018$.

Figure~\ref{fig:results-3-ele-cant}(b) shows the distribution of the relative error $\mathcal{R}$ obtained by varying the time-evolution parameter $t$ and the number of QSP phase factors $p$.
The overall trend is consistent with the tensile simulation: 
QGFA exhibits rapid convergence when the initial state vector $\bm{u}(0)$ is appropriately prepared, 
achieving small relative errors even for moderate values of $t$ and $p$.
However, due to the higher condition number $\kappa$, the relative errors in this problem are generally larger across the entire parameter range.
This result confirms that, although the convergence characteristics of QGFA remain stable, 
the achievable accuracy is influenced by the condition number of the stiffness matrix.

Figure~\ref{fig:results-qsp-func} illustrates the QSP approximation results for the target functions 
$g_1(s(x,\varepsilon))$, $\tilde{g}_2(s(x,\varepsilon))$, and $g_{\kappa,\epsilon}^{\text{inv}}(x)$, 
with polynomial degrees of $d=1600$, $1600$, and $3244$, respectively.
The time evolution parameter is set to $t=10\kappa$ in $g_1$ and $\tilde{g}_2$.
For both $g_1$ and $\tilde{g}_2$, the QSP responses exhibit good agreement with the target functions across the entire range of $x\in[0,1]$.  
The approximation error for $g_1$ is on the order of $10^{-2}$, while that for $\tilde{g}_2$ is further reduced to below $10^{-6}$.  
These results indicate that the even--function regularization introduced through the soft absolute function $s(x,\varepsilon)$ 
stabilizes the polynomial approximation throughout $x\in(0,1]$.
This regularization enables QSP to approximate exponentially decaying functions with high numerical stability.
A closer comparison between $g_1$ and $\tilde{g}_2$ reveals a difference in their sensitivity to approximation errors.  
Since $g_1(s(x,\varepsilon))$ rapidly decays toward zero, its approximation errors have only a minor effect on the overall solution.  
In contrast, $\tilde{g}_2(s(x,\varepsilon))$ contributes directly to the load-dependent term in the gradient--flow dynamics, and therefore requires higher precision to ensure accurate convergence.

Meanwhile, the QSP approximation of the inverse function $g_{\kappa,\epsilon}^{\text{inv}}(x)$ in Fig.~\ref{fig:results-qsp-func}(c) 
shows partially high accuracy, reaching absolute errors as low as $10^{-14}$ in mid-- to large--$x$ regions, but suffers from a clear loss of accuracy near small $x$ values close to $1/\kappa$.
This degradation originates from the intrinsic singularity of the $1/x$ form, where the function diverges as $x\to 0$.  
Although the degree $d=3244$ ensures smooth approximation across most of the spectrum, 
the accumulated error in the small--$x$ region affects the overall accuracy of the reconstructed solution vector.  
This observation highlights a fundamental advantage of QGFA over the conventional QMIA approach: 
the exponential-type functions in QGFA ($e^{-xt}$ and $(1-e^{-xt})/x$) are numerically well--behaved and far easier to approximate with QSP polynomials than the inverse function itself, leading to faster convergence and improved stability, especially for systems with large condition numbers.
\section{Conclusion}\label{sec:conclusion}

In this study, we propose QGFA, a novel quantum algorithm for solving SPD linear systems based on the time--evolution interpretation of variational principles.
Unlike conventional quantum linear solvers that directly approximate the matrix inverse, 
QGFA realizes the solution through the gradient--flow time evolution of the variational functional.
By combining the QET and LCU framework, the algorithm enables efficient quantum realization of the exponential matrix functions involved in the time evolution.

The proposed QGFA is demonstrated in the context of the FEM for solid mechanics problems under plane--stress conditions.
Through tensile and cantilever beam simulations, we confirmed that QGFA can accurately reproduce the classical FEM solutions.
The results show that the algorithm achieves rapid convergence for moderate time--evolution parameters $t$ and a small number of QSP phase factors $p$, particularly when an appropriate initial vector $\bm{u}(0)$, i.e., a hot-start solution, is provided.
Compared with QMIA, QGFA achieves smaller relative errors with fewer phase factors, 
demonstrating its potential as a numerically efficient alternative for SPD systems.
Furthermore, the accuracy and convergence characteristics are found to depend on the condition number of the stiffness matrix $\bm{K}$, suggesting possible strategies for preconditioning and adaptive parameter selection.

From a broader perspective, the QGFA framework can serve as a foundation for Quantum CAE.
It provides a physically interpretable bridge between classical iterative solvers and quantum computational paradigms.
Following the insights of Endo et al.~\cite{endo2025quantum} on the importance of pre-initialization in nonlinear quantum simulations and the preconditioning strategies by Raisuddin et al.~\cite{raisuddin2024qrls,raisuddin2024quantum}, QGFA may also be extended as a preconditioning approach for more complex, nonlinear, or multiphysics quantum simulations.
Future work will focus on integrating QGFA with quantum iterative refinement techniques 
and exploring its applications to nonlinear and time-dependent problems, 
paving the way toward fully realized Quantum CAE.

\appendix
\section{Soft Absolute Function}\label{sec:appendix-saf}

\begin{figure*}[t]
    \centering
    \includegraphics[width=0.9\linewidth]{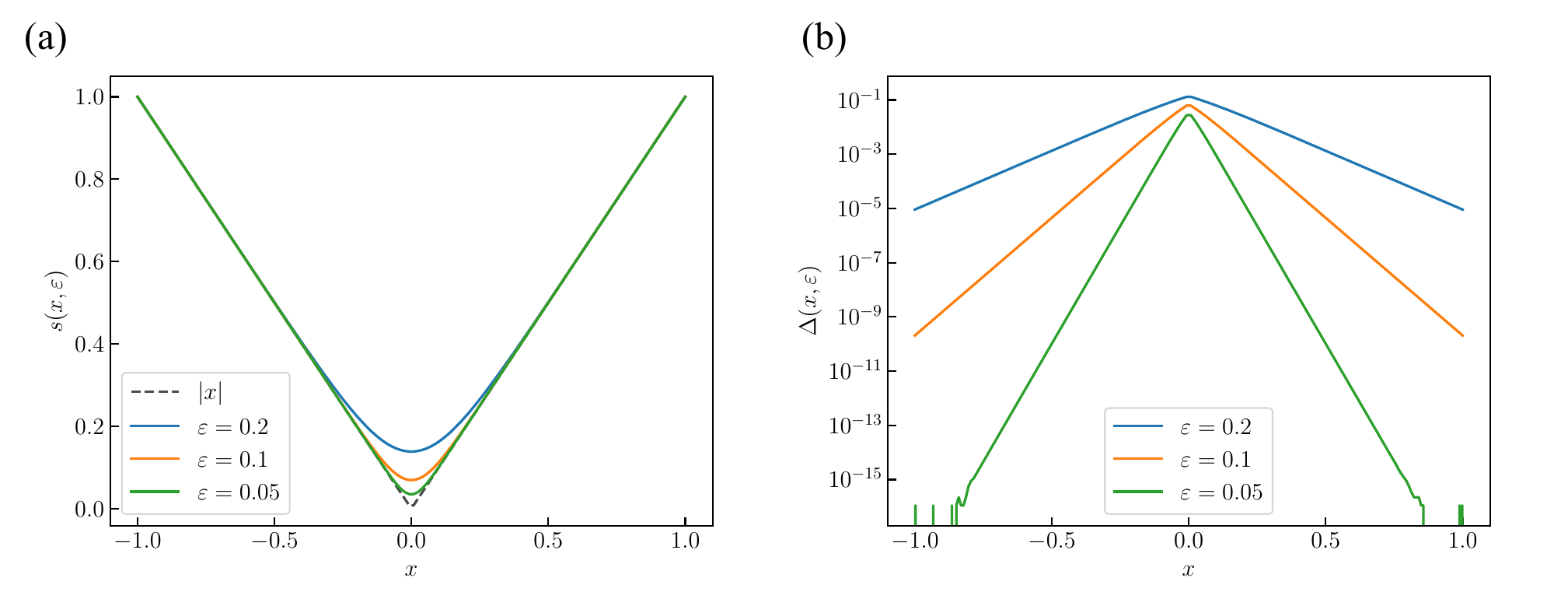}
    \caption{Properties of the soft absolute function. (a) Comparison between the soft absolute function $s(x,\varepsilon)$ and the standard absolute value $|x|$ for different values of the smoothing parameter $\varepsilon$. (b) The corresponding absolute approximation error $\Delta(x,\varepsilon)$ on a logarithmic scale, which decreases rapidly as $|x|/\varepsilon$ increases. The values for $\varepsilon=0.05$ exhibit round--off errors.}
    \label{fig:softabs-prop}
\end{figure*}

The absolute value function $|x|$ plays an important role in various mathematical operations. 
However, it is not an analytic function, and its steepness often causes difficulties in computations, such as in QSP \cite{motlagh2024generalized,dong2021efficient,low2017optimal} and Chebyshev expansion \cite{mason2002chebyshev,toh1998chebyshev,press1992numerical}.
To address this, we define the \emph{soft absolute function} $s(x,\varepsilon)$ as:
\begin{align}
    s(x,\varepsilon) 
    &= \varepsilon \ln \left( e^{\frac{x}{\varepsilon}} + e^{-\frac{x}{\varepsilon}} \right) \nonumber \\
    &= \varepsilon \ln \!\left( 2\cosh\!\left(\tfrac{x}{\varepsilon}\right) \right),
    \label{eq:softabs-app}
\end{align}
where $\varepsilon > 0$ is a smoothing parameter. 
Since $\cosh(z) > 0$ for all $z \in \mathbb{R}$, the soft absolute function $s(x,\varepsilon)$ is an analytic function of class $C^{\infty}$ on $\mathbb{R}$. 
This function provides a unique smooth approximation to the absolute value function with high accuracy.
Its first derivative is given as follows:
\begin{align}
  \frac{\partial s(x,\varepsilon)}{\partial x}
  &= \frac{\varepsilon}{2\cosh{\frac{x}{\varepsilon}}} \frac{1}{\varepsilon}\left(2\sinh{\frac{x}{\varepsilon}}\right) \nonumber\\
  &= \tanh\!\left(\frac{x}{\varepsilon}\right).
\end{align}
The second derivative is given as follows:
\begin{align}
  \frac{\partial^2 s(x,\varepsilon)}{\partial x^2}
  = \frac{1}{\varepsilon \cosh^2\!\left(\frac{x}{\varepsilon}\right)}
  = \frac{1}{\varepsilon}\operatorname{sech}^2\!\left(\frac{x}{\varepsilon}\right) \ge 0,
\end{align}
which shows that $s(x,\varepsilon)$ is convex.

Furthermore, from Eq.~\eqref{eq:softabs-app}, the absolute value function $|x|$ can be factored out as follows:
\begin{align}
  s(x,\varepsilon)
  = |x| + \varepsilon \ln\!\left(1 + e^{-\frac{2|x|}{\varepsilon}}\right).
  \label{eq:softabs-fac}
\end{align}
Therefore, by taking a smaller $\varepsilon$, the soft absolute function converges to the exact absolute value function:
\begin{align}
    \lim_{\varepsilon \to 0} s(x,\varepsilon) = |x|.
    \label{eq:aymp-softabs}
\end{align}
To evaluate the accuracy of $s(x,\varepsilon)$, we derive the approximation error $\Delta(x,\varepsilon)$ from Eq.~\eqref{eq:softabs-fac} as follows:
\begin{align}
    \Delta(x,\varepsilon)
    = s(x,\varepsilon)-|x|
    = \varepsilon\ln\left(1+e^{-\frac{2|x|}{\varepsilon}}\right)\geq 0\ .
    \label{eq:abserror-softabs}
\end{align}
Since $\ln(1+z) \le \ln 2$ for $z\in[0,1]$ and $z=e^{-2|x|/\varepsilon}\in(0,1]$, we obtain:
\begin{align}
    0 \leq \Delta(x,\varepsilon)\le \varepsilon\ln 2\xrightarrow[\varepsilon\to 0]{} 0\ ,
\end{align}
which shows that $s(x,\varepsilon)$ converges uniformly to $|x|$ on $\mathbb{R}$ as $\varepsilon\to 0$.
The first derivative of $\Delta(x,\varepsilon)$ with respect to $|x|$ is given as follows:
\begin{align}
    \frac{\partial\Delta(x,\varepsilon)}{\partial |x|}
    &= \frac{\varepsilon}{1+e^{-\frac{2|x|}{\varepsilon}}}\frac{\partial}{\partial |x|}\left(e^{-\frac{2|x|}{\varepsilon}}\right)\nonumber\\
    &= -\frac{2\,e^{-\frac{2|x|}{\varepsilon}}}{1+e^{-\frac{2|x|}{\varepsilon}}}<0\ ,
    \label{eq:deriv-error-softabs}
\end{align}
which implies that $\Delta(x,\varepsilon)$ is strictly decreasing in $|x|$ and attains its maximum at $x=0$.

When $|x|\gg\varepsilon$, the exponential factor $e^{-2|x|/\varepsilon}$ becomes very small. 
Therefore, in Eq.~\eqref{eq:abserror-softabs} the argument of the logarithm is close to $1$. 
For $z\to 0$, the Taylor expansion of the logarithm is:
\begin{align}
    \ln(1+z)=z-\tfrac{1}{2}z^2+O(z^3)\ .
\end{align}
By substituting $z=e^{-\frac{2|x|}{\varepsilon}}$, we obtain:
\begin{align}
    \Delta(x,\varepsilon)
    &= \varepsilon \left\{e^{-\frac{2|x|}{\varepsilon}}
    -\tfrac{1}{2}e^{-\frac{4|x|}{\varepsilon}}
    +O\left(e^{-\frac{6|x|}{\varepsilon}}\right)\right\}\nonumber \\
    &= \varepsilon e^{-\frac{2|x|}{\varepsilon}}
    + O\left(e^{-\frac{4|x|}{\varepsilon}}\right)\ .
    \label{eq:decay-softabs}
\end{align}
Therefore, the approximation error decays exponentially fast in $|x|/\varepsilon$. 
In particular, for $|x|\gg \varepsilon$, the leading--order term is $\varepsilon e^{-2|x|/\varepsilon}$, while all higher--order corrections are suppressed by additional exponential factors.

Here, the properties of $s(x,\varepsilon)$ are illustrated numerically in Fig.~\ref{fig:softabs-prop}. 
Figure~\ref{fig:softabs-prop}(a) shows that as $\varepsilon$ decreases, the soft absolute function $s(x,\varepsilon)$ approaches the absolute value function $|x|$. 
These results correspond to what is represented in Eq.~\eqref{eq:aymp-softabs}.
Figure~\ref{fig:softabs-prop}(b) demonstrates that the corresponding approximation error decays rapidly with increasing $|x|/\varepsilon$. 
As mentioned in Eq.~\eqref{eq:deriv-error-softabs} and \eqref{eq:decay-softabs}, the error is maximized at $x=0$, decreases monotonically as $|x|$ increases, and exhibits exponential decay in the regime $|x|\gg \varepsilon$. 
For $\varepsilon=0.05$, small oscillations appear around $10^{-15}$, which are attributed to numerical round--off errors.

\section{Formulation of FEM for Linear Elastic Deformation Analysis}\label{sec:appendix-fem}

Problems in solid mechanics can often be formulated as a linear system as follows:
\begin{align}
    \bm{K}^{\text{G}} \bm{u} = \bm{f}^{\text{G}}\,,
    \label{eq:stiffness-SPD-linear-equation}
\end{align}
where $\bm{K}^{\text{G}}$, $\bm{u}$, and $\bm{f}^{\text{G}}$ denote the global stiffness matrix, the displacement vector, and the global load vector, respectively.
By solving this linear equation, the displacement field over the entire domain is obtained, from which the stress field and other mechanical quantities can be computed in post-processing.

In this study, we focus on the displacement-based FEM, in which specific displacement boundary conditions $\bm{u}_{\text{D}}$ 
(Dirichlet boundary conditions) are prescribed on a portion of the boundary.
The remaining degrees of freedom, denoted by $\bm{u}_{\text{N}}$, correspond to the unknown displacements on the Neumann boundaries, 
which are determined by solving Eq.~\eqref{eq:stiffness-SPD-linear-equation}.
By separating the degrees of freedom associated with Dirichlet and Neumann boundaries, 
Eq.~\eqref{eq:stiffness-SPD-linear-equation} can be expressed in a block matrix form as follows:
\begin{align}
    \left(
    \begin{matrix}
        \bm{K}_{\text{DD}} & \bm{K}_{\text{DN}}\\
        \bm{K}_{\text{ND}} & \bm{K}_{\text{NN}}
    \end{matrix}
    \right)\left(
    \begin{matrix}
        \bm{u}_{\text{D}}\\ \bm{u}_{\text{N}}
    \end{matrix}
    \right)=\left(
    \begin{matrix}
        \bm{f}_{\text{D}}\\\bm{f}_{\text{N}}
    \end{matrix}
    \right)\,,
\end{align}
where $\bm{K}_{\text{DD}}$, $\bm{K}_{\text{DN}}$, $\bm{K}_{\text{ND}}$, and $\bm{K}_{\text{NN}}$ represent 
the submatrices of the global stiffness matrix corresponding to the Dirichlet and Neumann degrees of freedom, respectively.
Similarly, $\bm{f}_{\text{D}}$ and $\bm{f}_{\text{N}}$ represent the external load components applied to the Dirichlet and Neumann boundaries, respectively.
Note that $\bm{K}_{\text{DD}}$, $\bm{K}_{\text{DN}}$, $\bm{K}_{\text{ND}}$, $\bm{K}_{\text{NN}}$, $\bm{u}_{\text{D}}$, and $\bm{f}_{\text{N}}$ are given quantities. 
The unknown vectors $\bm{u}_{\text{N}}$ and $\bm{f}_{\text{D}}$ can be obtained as follows:
\begin{align}
    &\bm{u}_{\text{N}}=\bm{K}_{\text{NN}}^{-1} \left(\bm{f}_{\text{N}} - \bm{K}_{\text{ND}} \bm{u}_{\text{D}} \right)\,,\label{eq:neuamann-SPD-classic} \\
    &\bm{f}_{\text{D}} = \bm{K}_{\text{DD}} \bm{u}_{\text{D}} + \bm{K}_{\text{DN}} \bm{u}_{\text{N}}\,.
    \label{eq:fd}
\end{align}
Equation~\eqref{eq:fd} is evaluated after solving Eq.~\eqref{eq:neuamann-SPD-classic}. 
Therefore, our primary objective is to compute $\bm{u}_{\text{N}}$.
Eventually, the stiffness equation can be reformulated without explicitly extracting $\bm{u}_{\text{D}}$ as follows:
\begin{align}
    \left(
    \begin{matrix}
        \bm{I} & \bm{0}\\
        \bm{0} & \bm{K}_{\text{NN}}
    \end{matrix}
    \right)\left(
    \begin{matrix}
        \bm{u}_{\text{D}}\\ \bm{u}_{\text{N}}
    \end{matrix}
    \right)=\left(
    \begin{matrix}
        \bm{u}_{\text{D}}\\\bm{f}_{\text{N}}-\bm{K}_{\text{ND}}\cdot\bm{u}_{\text{D}}
    \end{matrix}
    \right)\,.
\end{align}
This representation maintains the original system size while directly enforcing the Dirichlet boundary conditions through the modified block structure. 
In the context of quantum computing, it is often difficult to extend or modify the dimension of a state vector once it has been initialized. 
Therefore, it is advantageous to formulate the problem in a unified vector representation that already includes the prescribed displacements $\bm{u}_{\text{D}}$ from the beginning.
This formulation can be directly incorporated into the QGFA framework, allowing the computation to proceed without any modification to the problem dimension as follows:
\begin{align}
    &\bm{K} = \left(
    \begin{matrix}
        \bm{I} & \bm{0}\\
        \bm{0} & \bm{K}_{\text{NN}}
    \end{matrix}
    \right)\,,\nonumber\\
    &\bm{f} = \left(
    \begin{matrix}
        \bm{u}_{\text{D}}\\\bm{f}_{\text{N}}-\bm{K}_{\text{ND}}\cdot\bm{u}_{\text{D}}
    \end{matrix}
    \right)\,,\label{eq:fem-input-qgfa}\\
    &\bm{u}(0) = \left(
    \begin{matrix}
        \bm{u}_{\text{D}}\\ \bm{0}
    \end{matrix}
    \right)\,.\nonumber
\end{align}
Here, $\bm{K}$ is SPD matrix, and the initial vector $\bm{u}(0)$ provides a hot start for the optimization process, which can also be beneficial from the viewpoint of variational optimization.

\section{Quantum Matrix Inverse Algorithm}\label{sec:appendix-inv}

In this section, we provide an overview of the QMIA.  
To obtain the matrix inverse solution as expressed in Eq.~\eqref{eq:invref}, 
Gily{\'e}n et al.~\cite{gilyen2019quantum} proposed an approximate inverse function $g^{\text{inv}}_{\kappa,\epsilon}(x)$ as follows:
\begin{align}
    g^{\text{inv}}_{\kappa,\epsilon}(x) = \frac{1-\left( 1 - x^2 \right)^{\kappa^{2}\log\left(\kappa/\epsilon\right)}}{x}\,,
    \label{eq:inv-qsp}
\end{align}
where $\epsilon$ represents the approximation threshold.
The polynomial degree is determined as the function of $\kappa$ and $\varepsilon$, which is $d(\kappa,\epsilon) = \sqrt{\kappa^{2}\log\left(\kappa/\epsilon\right)\log\left\{ 4 \kappa^{2}\log\left(\kappa/\epsilon\right) / \epsilon \right\}}$~\cite{martyn2021grand,gilyen2019quantum}.

We employ the \texttt{pyQSP} package~\cite{martyn2021grand,gilyen2019quantum,dong2021efficient,chao2020finding,haah2019product} 
to determine the phase factors 
$\bm{\Phi}^{\text{inv}} = (\phi^{\text{inv}}_0,\phi^{\text{inv}}_1,\cdots,\phi^{\text{inv}}_{d-1}) \in \mathbb{R}^{d}$ 
corresponding to Eq.~\eqref{eq:inv-qsp}.  
The associated unitary matrix can then be constructed as follows:

\begin{align}
    \bm{U}_{\bm{\Phi}^{\text{inv}}} 
    &= \bm{R}_{z}\left(\phi^{\text{inv}}_{0}\right) \prod_{j=1}^{(d-1)/2} \left[\widetilde{\bm{U}}^{\dagger}_{\bm{K}} \bm{R}_{z}\left(\phi_{2j-1}^{\text{inv}}\right)\widetilde{\bm{U}}_{\bm{K}} \bm{R}_{z}\left(\phi_{2j}^{\text{inv}}\right)\right] \nonumber\\
    &= \left(
    \begin{matrix}
        \text{P}_{\text{QET}}\left[g_{\kappa,\epsilon}^{\text{inv}}\left( \tilde{\bm{K}} \right)\right]_{d-1}  & * \\
        * & *
    \end{matrix}
    \right)\,,\label{eq:QETinv}
\end{align}
Note that Eqs.~\eqref{eq:QETinv} corresponds to the standard QET construction when $d$ is odd.
The quantum solution vector of QMIA $\bm{u}_{\text{qc}}^{\text{inv}}$ can be represented in the state--vector form as:
\begin{align}
    \ket{\bm{u}_{\text{qc}}^{\text{inv}}} = \bm{U}_{\bm{\Phi}^{\text{inv}}} \ket{\bm{f}}\,.
\end{align}
Ideally, the resulting solution is extracted as follows:
\begin{align}
    \bm{u}_{\text{qc}}^{\text{inv}} = \frac{\text{P}_{\text{QET}}\left[g_{\kappa,\epsilon}^{\text{inv}}\left( \tilde{\bm{K}} \right)\right]_{d-1} }{||\bm{K}||} \bm{f}\,.
\end{align}

\section*{Acknowledgment}

The authors would like to sincerely thank Dr.~Yasunari Suzuki from the Center for Quantum Computing at RIKEN for providing us with insightful lectures on quantum linear system algorithms in the fault-tolerant quantum computing (FTQC) era. 
This work was performed for Council for Science, Technology and Innovation (CSTI), Cross--ministerial Strategic Innovation Promotion Program (SIP), ``Promoting the application of advanced quantum technology platforms to social issues'' (Funding agency : QST).

\bibliography{apssamp}% Produces the bibliography via BibTeX.

% The \nocite command causes all entries in a bibliography to be printed out
% whether or not they are actually referenced in the text. This is appropriate
% for the sample file to show the different styles of references, but authors
% most likely will not want to use it.
% \nocite{*}

%\bibliography{apssamp}% Produces the bibliography via BibTeX.
%\clearpage

\end{document}